# Isotope effect on radiative thermal transport

Lanyi Xie[1,3] and Bai Song[1,2,3†]

[1]*Department of Energy and Resources Engineering, Peking University, Beijing 100871, China.*

[2]*Department of Advanced Manufacturing and Robotics, Peking University, Beijing 100871, China.*

[3]*Beijing Innovation Center for Engineering Science and Advanced Technology, Peking University, Beijing 100871, China.*

[†]Corresponding author. Email: songbai@pku.edu.cn

**ABSTRACT**

Isotope effects on heat conduction and convection have been known for decades. However, whether thermal radiation can be isotopically engineered remains an open question. Here, we predict over 3-orders-of-magnitude variation of radiative heat flow with varying isotopic compositions for polar dielectrics at room temperature. We reveal this as an isotope mass effect which induce phonon line shift and broadening that in turn affect phonon-mediated resonant absorption both in the near and far field. In contrast, the isotope effect is negligible for metals and doped semiconductors which largely depend on free carriers. We also discuss the role of temperature with regard to surface mode excitation.

Many elements have two or more stable isotopes which differ in their atomic mass and sometimes also in the nuclear spin [1]. Materials with different isotopic compositions often exhibit distinct physical properties, especially those closely related to the crystal structures and lattice dynamics as a result of various isotope mass effects [2-5]. As an example, enhancement of phonon thermal conductivity via isotopic enrichment have been predicted and observed since the 1950s [6-16], considering the reduced scattering of phonons due to smaller atomic mass fluctuations. With continued progress in the growth of isotope-controlled crystals, substantial isotope effects of over 90% and 150% have recently been measured in bulk cubic boron nitride (cBN) [15] and silicon (Si) nanowires [16], respectively. Further, isotope effects on fluid viscosity have also long been appreciated which is key to convective heat transfer [17-19]. Despite decades of research on conduction and convection which represent two of the three basic modes of heat transfer [20], isotope effects on the third mode, i.e., thermal radiation, remains to be explored and is the focus of our present work.

Thermal radiation both in the far and near field can be consistently described with fluctuational electrodynamics [21]. Key to this theoretical approach is the complex permittivities of the interacting bodies. For metals and doped semiconductors, the permittivity is essentially determined by the free carriers, especially the carrier density, and is given by the Drude model [22]. For polar dielectrics, the permittivity is usually captured with the Lorentz model [22] which is dictated by the zone-center optical phonons. Both the carrier density and phonon frequency depend crucially on the lattice constant of the crystal [23]. The variation of lattice constant with isotopic composition has been extensively studied using *x*-ray diffraction and the *x*-ray standing wave technique, and is viewed as a manifestation of the zero-point energy [24-26]. Similarly, isotope-induced phonon line shift and broadening have been widely observed via

Raman spectroscopy [27-29], neutron scattering [30], and electron energy loss spectroscopy [31,32]. In addition, first-principles calculations have also been adopted to study isotope-dependent lattice dynamics [12,33-35]. These results form the basis of our theoretical inquiry into the isotopic dependence of thermal radiation.

Here, we consider the basic case of radiative thermal transport between two parallel planes with a focus on materials supporting electromagnetic surface modes [36,37]. We begin with cBN which is a prototypical polar dielectric. For bulk cBN at room temperature, we show that the radiative heat flow can vary by over 10-fold with varying boron isotope ratio. With thin films instead, heat-flow variations up to 1650 are predicted. This isotope effect is largely attributed to the match and mismatch of the narrow surface phonon polariton (SPhP) peaks of cBN which shift with its reduced atomic mass. Subsequently, we consider isotopically engineered lithium hydride which offer some of the largest relative mass variations, and accordingly predict over 7260-fold heat-flow variations. However, with increasing damping of the SPhPs, the isotope effect is suppressed by orders of magnitude. In addition, we examine the isotope effects for elemental metals and doped semiconductors with lithium, copper, and silicon as examples, obtaining values on the order of $10^{-4}$. This negligibly small isotopic dependence originates from the limited thermal excitation and more importantly strong damping of surface plasmon polaritons. To conclude, we systematically illustrate the impact of isotope-induced phonon line shift and broadening by using a hypothetical polar dielectric. The influence of temperature is also discussed.

Thermal radiation between polar dielectrics is of great importance especially in the near field due to the presence of surface phonon polaritons (SPhPs) [38-44]. The permittivity of polar dielectrics is given by the Lorentz model [22] as

$$\varepsilon(\omega) = \varepsilon_\infty \left(1 - \frac{\omega_{\text{LO}}^2 - \omega_{\text{TO}}^2}{\omega^2 - \omega_{\text{TO}}^2 + i\Gamma\omega}\right), \tag{1}$$

where $\varepsilon_\infty$ is the high-frequency dielectric constant, $\omega_{\text{TO}}$ and $\omega_{\text{LO}}$ are respectively the frequencies of the transverse and longitudinal optical phonons at the center of the Brillouin zone, and $\Gamma$ is the damping factor related to the phonon linewidths. Usually, $\varepsilon_\infty$ is insensitive to isotopes because it mainly depends on the polarizability of the valence electrons [45], while the phonon frequencies and the damping factor are all sensitive to the isotopic composition.

To start with, we consider cBN. Natural nitrogen consists of 99.6% $^{14}$N and is treated as isotopically pure for simplicity. In contrast, natural boron comprises of 19.9% $^{10}$B and 80.1% $^{11}$B, and often leads to substantial isotope effects [15,46-49]. In Fig. 1(a), we plot the $\omega_{\text{TO}}$ and $\omega_{\text{LO}}$ of cBN which increase roughly linearly with the fraction of $^{10}$B from 1052 cm$^{-1}$ to 1081 cm$^{-1}$ and from 1280 cm$^{-1}$ to 1316 cm$^{-1}$, respectively. The lines are calculated using a spring-mass oscillator model with $\omega \propto \mu^{-1/2}$ which agree well with previous Raman measurements (symbols) [15]. Here, $\mu$ denotes the reduced mass of isotopically engineered cBN and varies with the average atomic mass of boron. In addition, we consider the phonon linewidth which usually increases with mass fluctuation [Fig. 1(a)] due to phonon-isotope scattering and provides an isotope-dependent $\Gamma$ [10,12]. For high-quality cBN crystals, both the Raman measured and first-principles calculated phonon linewidths are close to 1 cm$^{-1}$ [15]. To study the isotope effect on thermal radiation, we employ cBN with three representative isotope ratios: c$^{11}$BN with pure $^{11}$B, c$^{eq}$BN with 50% $^{10}$B, and c$^{10}$BN with pure $^{10}$B. The calculated $\omega_{\text{TO}}$, $\omega_{\text{LO}}$, and $\Gamma$ are listed in Table S1 of the Supplemental Material [50], along with those of natural cBN. Substantial frequency shifts are observed in the corresponding permittivities, together with some peak broadening [Fig. 1(b)].

For two parallel plates separated by a vacuum gap, the total radiative heat transfer coefficient (HTC) in the linear regime can be written in a Landauer-like form as [21]

$$h(T,d) = \int_0^\infty h_\omega d\omega = \int_0^\infty \frac{d\omega}{4\pi^2} \frac{\partial \Theta(\omega,T)}{\partial T} \int_0^\infty dk k [\tau_s(\omega,k) + \tau_p(\omega,k)]. \quad (2)$$

Here, $h_\omega$ is the spectral HTC (sHTC), $d$ is the gap size, $T$ is the absolute temperature, $\Theta(\omega,T) = \frac{\hbar\omega}{\exp(\hbar\omega/k_B T)-1}$ is the mean energy of a harmonic oscillator less the zero-point contribution, $\omega$ is the angular frequency, $k$ is the parallel wavevector, $\tau_s$ and $\tau_p$ are respectively the transmission probabilities of the *s*- and *p*-polarized modes which depend closely on the permittivities (see [50]).

In Fig. 2(a), we plot the total HTCs versus the gap size between bulk cBN plates at 300 K for four representative configurations pairing respectively $c^{11}BN$ with $c^{11}BN$, $c^{10}BN$ with $c^{10}BN$, $c^{eq}BN$ with $c^{eq}BN$, and $c^{11}BN$ with $c^{10}BN$. Compared to the asymmetric case of $c^{11}BN$-$c^{10}BN$, the three symmetric pairs feature much larger HTCs at small gaps. This difference originates from the match and mismatch of the SPhPs on the two plates as revealed by the corresponding sHTCs and the electromagnetic local density of states (LDOS) [38] shown in Fig. 2(b)-(c). Moreover, $c^{eq}BN$-$c^{eq}BN$ has the highest HTC among the three symmetric pairs because the larger damping factor (1.3 cm$^{-1}$ for $c^{eq}BN$ versus ~0.5 cm$^{-1}$ for $c^{11}BN$ and $c^{10}BN$) results in a broader SPhP peak. Together, these cases highlight the two basic mechanisms underlying the isotope dependence of near-field radiative thermal transport between polar dielectrics, that is, the frequency shift and line broadening of the zone-center optical phonons.

To quantify the isotope effect, we first focus on the impact of phonon frequency shift by fixing one plate as $c^{11}BN$ and varying the $^{10}B$ fraction of the other denoted as $c^{eng}BN$. The magnitude of the isotope effect is defined as $\eta = h_{max}/h_{min} - 1$ where $h_{\max}$ and $h_{\min}$ are the maximum and minimum HTC, respectively. As expected, $h_{\min}$ corresponds to the case of $c^{11}BN$-$c^{10}BN$ which features the largest SPhP mismatch. Meanwhile, $h_{\max}$ is obtained with $c^{11}BN$-$c^{11}BN$. With bulk cBN, we obtain $\eta$ on the order of 10 at nanometer gaps [Fig. 2(a) inset]. In the far field, however, $\eta$ is negligibly small (~0.01) due to the vanishing contribution of SPhPs.

The predominant role of the SPhPs motivates us to further explore the possibility of engineering the isotope effects using, for example, thin films [41,51-53]. In Fig. 3(a), we show the calculated $\eta$ for cBN films of varying thickness *t* from 100 μm to 1 nm, at three representative gap sizes of 100 nm, 1 μm, and 100 μm. In general, $\eta$ increases with decreasing film thickness, although small fluctuations appear for films of intermediate thickness [50]. With nanometer-thin films, the isotope effect is enhanced by up to 2-orders-of-magnitude in the near field to over 1000. More intriguingly, nanofilms also lead to a dramatically enhanced $\eta$ of over 100 in the far field.

To understand the mechanisms underlying the thin-film enhancement of isotope effect both in the near and far field, we plot in Fig. 3 the sHTCs for the case of 1-nm-thick cBN films at the gap sizes of 100 nm and 100 μm, together with the corresponding LDOS. The sHTCs for bulk cBN are also shown for comparison. At 100 nm-gap, two characteristics in the sHTCs stand out [Fig. 3(b)]. First, the number of peaks double for the nanofilm because the SPhPs at the two interfaces couple within the film and split into a low-frequency symmetric branch plus a high-frequency antisymmetric branch [51], as confirmed by the LDOS [Fig. 3(c)]. More importantly, the non-SPhP contribution, especially from the frustrated modes, drops by orders of magnitude with nanofilms due to the reduced volume [52]. As a result, the total HTCs in the nanofilm cases become dramatically more sensitive to the SPhPs which in turn are dictated by the isotope-engineered optical phonons. This explains the substantially increased isotope effect for the thin-film geometry in the near field.

In the far field, however, SPhPs barely contribute. Indeed, the sHTCs for bulk cBN is broadband both in the symmetric and asymmetric case [Fig. 3(d)]. Interestingly, as we go to nanofilms, the sHTCs clearly feature sharp peaks around $\omega_{TO}$ and $\omega_{LO}$ of $c^{11}BN$ and $c^{10}BN$ where LDOS peaks are also observed [Fig. 3(e)]. These peaks are orders-of-magnitude higher than the surrounding regions, analogous to the SPhP peaks in the near field. The reason for the far-field spectral peaks lies in the resonant absorption of infrared photons by the zone-center optical phonons. Briefly, the imaginary part of the permittivity reaches a maximum around $\omega_{TO}$ [Fig. 1(b)] which leads to strong absorption of both *s*- and *p*-polarized waves regardless of the incident angle. Meanwhile, the near-zero permittivity around $\omega_{LO}$ results in large absorption of obliquely incident *p*-polarized waves which is characteristic of the Berreman leaky modes [54-56]. As we go from bulk cBN to nanofilms, the absorption in most spectral regions drops dramatically while the phonon-induced resonant peaks remain strong. Therefore, their match and mismatch as a result of varying isotopic compositions dictates the heat flow, which explains the significant isotope effect in the far field.

In addition to phonon line shift, the damping factor may also have a large impact on the isotope effect because it affects the widths of the SPhP peaks. In light of the thin-film enhancement, we focus below only on the case of 1-nm-thick films. In Fig. 4(a), we plot $\eta$ for cBN as a function of gap size for three damping factors including ~0.5 cm$^{-1}$, 2 cm$^{-1}$, and 10 cm$^{-1}$, which represent crystals of different quality [22]. With $\Gamma \approx$ 0.5 cm$^{-1}$, $\eta$ goes up to 1650 at a 615 nm-gap; while with $\Gamma = 10$ cm$^{-1}$, a maximum $\eta$ of 9 is achieved at a 250 nm-gap. Overall, over 2-orders-of-magnitude suppression of $\eta$ is predicted with increasing $\Gamma$. This is because for the same line shift, wider SPhP peaks may experience smaller spectral mismatch, as shown in Fig. 4(b).

Inspired by the large isotope effect in cBN and enabled by the physical insight gained from it, we proceed to lithium hydride (LiX with Li = $^6$Li, $^7$Li and X = H, D, T) which is also a polar dielectric but contains two of the lightest elements and allows a very large relative mass variation (from $^6$LiH to $^7$LiT). Analogous to cBN, we compare the symmetric pair of $^7$LiT with the asymmetric pair of $^6$LiH and $^7$LiT. Their phonon properties are obtained from the literature and listed in Table S1 [50]. The calculated isotope effects for LiX are also shown in Fig. 4(a). Comparing $\Gamma$ = 2 cm$^{-1}$, 10 cm$^{-1}$, and $0.1\omega_{\text{TO}}$ (~37.9 cm$^{-1}$ for $^7$LiT and 59.5 cm$^{-1}$ for $^6$LiH) [57], 3-orders-of-magnitude suppression of $\eta$ with increasing $\Gamma$ is predicted, similar to the case of cBN. Remarkably, with $\Gamma$ = 2 cm$^{-1}$, $\eta$ reaches up to 7260 at a 720 nm-gap, which is about 40× the maximum $\eta$ for cBN. This illustrates the impact of the relative phonon line shift which scales as $\Delta\omega_{\text{LO}}/\omega_{\text{LO}} \approx \Delta(\mu^{-1/2})/\mu^{-1/2}$ and is about 2.8% for cBN and 56.4% for LiX. The large value for LiX is due to a large $\Delta\omega_{\text{LO}}$ combined with a small $\omega_{\text{LO}}$ (708 cm$^{-1}$ for $^7$LiT). The gap-dependence of $\eta$ is complex, although large values often appear at relatively large gaps (100s of nm).

In addition to polar dielectrics, we also explore the isotope effect for materials supporting surface plasmon polaritons (SPPs) such as metals and doped semiconductors, the permittivity of which is usually described by the Drude model as [22]

$$\varepsilon(\omega) = \varepsilon_\infty - \frac{\omega_p^2}{\omega^2 + i\Gamma\omega}. \qquad (3)$$

Here, $\omega_p = \sqrt{\frac{4\pi e^2}{m_e}n}$ is the plasma frequency, $e$ and $m_e$ are respectively the electron charge and effective mass, and $n$ is the number density which is inversely proportional to the volume per atom and thus cube of the lattice constant $a$ [23]. Isotopic dependence of the lattice constant has been widely studied. At relatively low temperatures, $a$ roughly varies with the reduced mass $\mu$ as $\Delta a/a = C\,\Delta\mu/\mu$, where $C$ is a material-specific coefficient on the order of $10^{-3}$ [58,59], which leads to an isotope-dependent $\omega_p$ since $\Delta\omega_p/\omega_p \approx 1.5\Delta a/a$ holds by definition. The lattice constants of various isotope-engineered elemental metal and semiconductor crystals are readily available in the literature, yielding $\Delta\omega_p/\omega_p$ on the order of $10^{-4}$, which is orders-of-magnitude smaller than $\Delta\omega_{\mathrm{LO}}/\omega_{\mathrm{LO}}$ for polar dielectrics.

With Li, copper (Cu) and doped-silicon ($d$-Si) as examples [23,60,61], we show the isotope effects as a function of gap size in Fig. 4(c), which remain on the order of $10^{-4}$. To be formally consistent with the polar dielectrics, we compare the symmetric pairs of $^{7}$Li, $^{65}$Cu, and $^{30}$Si with the asymmetric pairs of $^{6}$Li and $^{7}$Li, $^{63}$Cu and $^{65}$Cu, and $^{28}$Si and $^{30}$Si, respectively. The remarkably small $\eta$ arises for multiple reasons. First, $\Delta\omega_p/\omega_p$ is extremely small. Further, the SPPs for metals cannot be thermally excited at room temperature, in contrast to the SPhPs for polar dielectrics. More importantly, the damping factors are orders-of-magnitude larger in metals and semiconductors, leading to broad radiation spectra that suppress the isotope effect. Note that without narrow resonant peaks dominating the spectra, the sign of $\eta$ is less intuitive.

Now that we have identified the key mechanisms and material properties underlying the large isotope effect for real polar dielectrics, we proceed to a parametric study using hypothetical materials. Without loss of generality, we assume the $\varepsilon_\infty$, $\omega_{\mathrm{TO}}$, and $\omega_{\mathrm{LO}}$ of cBN, and vary $\Delta\omega_{\mathrm{LO}}/\omega_{\mathrm{LO}}$ from 1% to 100% and $\Gamma$ from 0.5 cm$^{-1}$ to 100 cm$^{-1}$. The calculated $\eta$ at $d = 615$ nm as a function of $\Delta\omega_{\mathrm{LO}}/\omega_{\mathrm{LO}}$ and $\Gamma$ at room temperature is mapped in Fig. 5(a), where the values for cBN crystals of different quality are marked. As expected, $\eta$ increases monotonically with decreasing damping and increasing line shift, spanning multiple orders of magnitude from $10^{-2}$ up to $10^6$. Nevertheless, a sharp dip is clearly visible at $\Delta\omega_{LO}/\omega_{LO} \approx 22\%$. This corresponds to the scenario where the line shift approaches the width of the Reststrahlen band, so that $\omega_{TO}$ of the lighter isotope combination matches $\omega_{LO}$ of the heavier.

In light of the essential role of surface waves, we further investigate the temperature dependence of the isotope effect with regard to mode excitation. Briefly, we calculate $\eta$ as a function of $T$ from 10 K to 1000 K for LiX, cBN, and two hypothetical polar dielectrics with $\omega_{LO} = 2000$ cm$^{-1}$ and 8000 cm$^{-1}$ which qualitatively represent $d$-Si of different doping concentrations given the similarity between the Lorentz and the Drude model (Table S3) [50]. For better comparison, $\Gamma = 2$ cm$^{-1}$ is used for all four materials and $\Delta\omega_{LO}/\omega_{LO}$ of cBN is also used for the hypothetical ones. The gaps correspond to maximum room-temperature $\eta$ in each case. Notably, the isotope effects always plateau at sufficiently low and high temperatures, with a sharp transition in between. The plateaus are dominated by the broadband frustrated modes and the narrow SPhPs at low and high $T$, respectively. Near the transition region, more and more SPhPs quickly get excited as $T$ increases, due to the exponential behavior of $\frac{\partial\Theta(\omega,T)}{\partial T}$. With increasing $\omega_{LO}$, the transition temperature increases from about 30 K for LiX to 60 K for cBN, and 120 K and 400 K for the materials imitating $d$-Si. The large $\eta$ for the latter again shows the impact of damping when compared to Fig. 4(c).

In summary, we have theoretically explored the isotopic dependence of radiative thermal transport focusing on materials supporting electromagnetic surface modes. The calculated isotope effects vary from over $10^3$ for polar dielectrics to $10^{-4}$ for metals and semiconductors, and are analyzed in terms of the frequency shift, damping factor, and thermal excitation of SPhPs and SPPs. A narrow radiation spectrum is at the heart of a large isotope effect. Thin films provide substantial enhancement over bulk materials. We reveal the underlying mechanism as an isotope mass effect which impacts the lattice constant and dynamics. For the future, it may be interesting to see if there is an isotope spin effect [62]. Our work highlights isotope engineering as a promising avenue for controlling thermal radiation. Similar effects are expected for the Casimir force which also arises from the fluctuating electromagnetic fields [59].

## ACKNOWLEDGMENT

This work was supported by the National Natural Science Foundation of China (Grant No. 52076002), the Beijing Innovation Center for Engineering Science and Advanced Technology, the XPLORER PRIZE from the Tencent Foundation, and the High-performance Computing Platform of Peking University. We thank Qizhang Li for verifying some of the calculations and Fuwei Yang for helpful discussions.

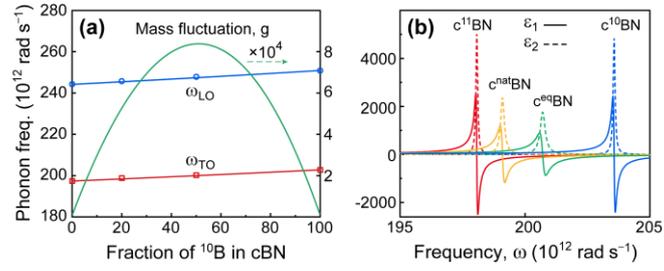

**FIG. 1. Isotope effects on the phonon properties and permittivity of cBN.** (a) Modeled (red and blue lines) and measured (symbols) frequency shift of the zone-center optical phonons with varying fraction of $^{10}$B, and the mass-fluctuation parameter $g$ (green line). (b) Permittivities of representative cBN. $\varepsilon_1$ and $\varepsilon_2$ are respectively the real and imaginary part.

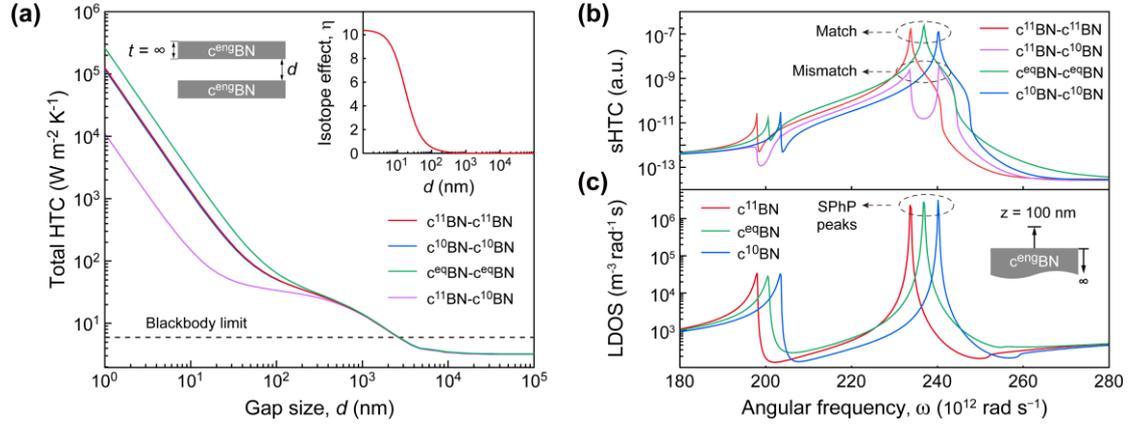

**FIG. 2. Isotope effects on radiative thermal transport between bulk cBN at room temperature.** (a) Total HTCs of representative pairs of isotope-engineered cBN as a function of gap size. Inset shows the magnitude of the isotope effect. (b) Spectral HTCs and (c) the corresponding LDOS with the SPhP peaks marked.

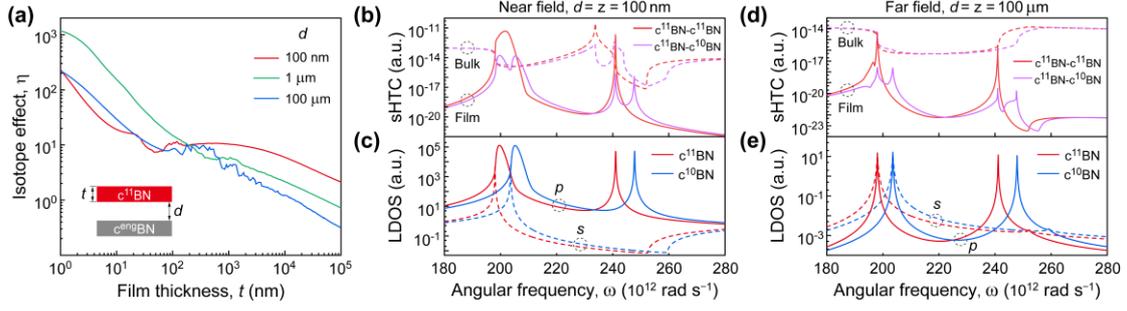

**FIG. 3. Thin-film enhancement of the isotope effect.** (a) Isotope effect at three typical gap sizes with varying cBN film thickness. (b) and (d) Spectral HTCs of the symmetric and asymmetric pairs in the near and far field, respectively. (c) and (e) The corresponding LDOS showing both the *s* and *p* polarizations.

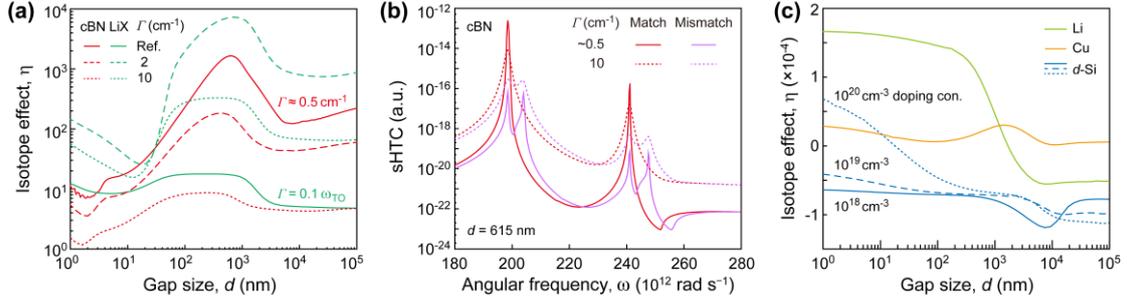

**FIG. 4. Isotope effect for real materials supporting surface waves.** (a) Polar dielectrics of varying damping factors. (b) Spectral HTC showing broadened SPhP peaks due to large damping. (c) Metals and semiconductors of different doping levels. All films are 1 nm thick.

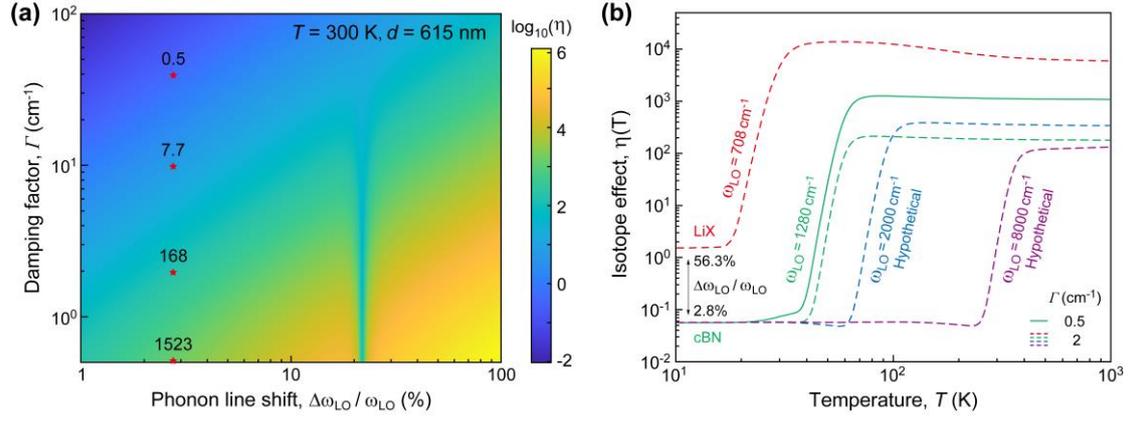

**FIG. 5. Parametric study of the isotope effect.** (a) A map with respect to the damping factor and the relative line shift. The $\varepsilon_\infty$, $\omega_{\mathrm{TO}}$, and $\omega_{\mathrm{LO}}$ of cBN are assumed. Select values are marked for reference at $\Delta\omega_{\mathrm{LO}}/\omega_{\mathrm{LO}} = 2.8\%$. (b) Temperature dependence of the isotope effect for materials with different $\omega_{\mathrm{LO}}$. Detailed properties are listed in Table S3 [50]. The gap sizes are all optimized to yield maximum $\eta$ at 300 K.